\documentstyle[draft,aps]{revtex}

\begin{document}
\title{Tripartite GHZ state generation with trapped ion in an optical cavity}
\author{S. Shelly Sharma$^{1,2}$}
\address{(1) Depto. de Fisica, Universidade Estadual de Londrina,\\
86040-370 Londrina, PR Brazil \\
(2) Instituto de Fisica ``GlebWataghin'',Universidade Estadual de Campinas, 
\\
13083-970 Campinas SP Brazil}
\maketitle

\begin{abstract}
We present a single step scheme to generate, experimentally, maximally
entangled tripartite GHZ state in a single ion, using trapped ion
interacting simultaneously with a resonant external laser field and red
sideband tuned quantized cavity field. Besides the simplicity of execution
and short operation time, GHZ state generation is reduction free.
\end{abstract}


Recent interest in tripartite entanglement is motivated by possible use in
quantum information processing. Tripartite entangled states have been shown
to be advantageous in comparison with bipartite Bell states in quantum
teleportation \cite{tele} and quantum dense coding \cite{code}. In
particular, three qubit Greenberger-Horne-Zeilinger (GHZ) \cite{ghz89} state
is found to be the maximally entangled state \cite{gisi98} in the sense that
it violates Bell inequalities maximally. GHZ state is unique in the sense
that two qubit EPR (Einstein-Podolsky-Rosen) state between any chosen pair
of qubits can be obtained from it. However, reversible generation of GHZ
from pairwise distribution of EPR states among three parties is not possible 
\cite{lind99}. Experimental efforts to realize multipartite entanglement
include photon polarization experiments \cite{ppol99}, Nuclear Magnetic
resonance \cite{nmr}, atoms-cavity experiments \cite{raus99}, trapped ions 
\cite{sack00} and more recently ions trapped inside an optical cavity \cite
{walt101,mund02}. For controlled information processing, cold ions in a
linear trap \cite{wine98} offer a promising approach, as each ion allows two
qubit state manipulation. With the ion trap placed inside a high finesse
optical cavity, we have at hand a tripartite system with additional control
mechanism offered by quantized cavity field. In this letter a single step
scheme to generate three qubit maximally entangled GHZ state, using trapped
ion interacting simultaneously with a resonant external laser and sideband
tuned quantized cavity field, is presented.

Consider a two level ion radiated by the single mode cavity field of
frequency $\omega _{c}$ and an external laser field of frequency $\omega _L$%
. Interaction with external laser field as well as the cavity field,
generates entanglement of internal states of the ion, vibrational states of
ionic center of mass motion and the cavity field state. The Hamiltonian, for
the case when center of the trap is close to the node of cavity field
standing wave, is given by 
\begin{eqnarray}
\hat{H} &=&\hat{H}_{0}+\hat{H}_{int},  \label{eq1} \\
\hat{H}_{0} &=&\hbar \nu \left( \hat{a}^{\dagger }\hat{a}+\frac{1}{2}\right)
+\hbar \omega _{c}\hat{b}^{\dagger }\hat{b}+\frac{\hbar \omega _{0}}{2}%
\sigma _{z},  \label{eq2}
\end{eqnarray}
\begin{eqnarray}
\hat{H}_{int} &=&\hbar \Omega \left[ \sigma _{+}\exp \left[ i\eta _{L}(\hat{a%
}^{\dagger }+\hat{a})-i\omega _{L}t\right] +h.c.\right]  \nonumber \\
&&+\hbar g(\sigma _{+}+\sigma _{-})\left( \hat{b}^{\dagger }+\hat{b}\right)
\sin \left[ \eta _{c}(\hat{a}^{\dagger }+\hat{a})\right] ,  \label{eq3}
\end{eqnarray}
where $\hat{a}^{\dagger }(\hat{a})$ and $\hat{b}^{\dagger }(\hat{b})$ are
creation(destruction) operators for vibrational phonon and cavity field
photon respectively and $\omega _{0}$ the transition frequency of the
two-level ion. The ion-phonon and ion-cavity coupling constants are $\Omega $
and $g$, whereas $\sigma _{k}(k=z,+,-)$ are the Pauli operators qualifying
the internal state of the ion. The Lamb-Dicke (LD) parameters relative to
the laser field and the cavity field are denoted by $\eta _{L}$ and $\eta
_{c}$ respectively.

The interaction picture Hamiltonian, determined by unitary transformation $%
U_{0}(t)=\exp \left[ -\frac{i\hat{H_{0}}t}{\hbar }\right] $ , is given in a
detailed form in Eq. (4) of ref. \cite{shel02}. Consider the ion interacting
simultaneously with a resonant external laser field of frequency $\omega
_{L}=\omega _{0}$ and red sideband tuned quantized cavity field, $\omega
_{0}-\omega _{c}=\nu .$ In rotating wave approximation, the relevant part of
the interaction picture Hamiltonian is 
\begin{eqnarray}
\hat{H_{I}} &=&\hbar \Omega \lbrack \sigma _{+}{\hat{O}_{0}^{L}}+\sigma _{-}{%
\hat{O}_{0}^{L}}]  \nonumber \\
&&+\hbar g\left[ \sigma _{+}\hat{b}{\eta _{c}}{\hat{O}_{1}^{c}}\hat{a}+h.c.%
\right] ,  \label{eq4}
\end{eqnarray}
where 
\begin{equation}
{{\hat{O}}_{k}}=\exp \left( -\frac{\eta ^{2}}{2}\right) \sum_{p=0}^{\infty }%
\frac{(i\eta )^{2p}\hat{a}^{\dagger p}\hat{a}^{p}}{p!\left( p+k\right) !}.
\label{eq5}
\end{equation}
The matrix element of diagonal operator ${{\hat{O}}_{k}}$ for a given
vibrational state $m$ is given by 
\begin{equation}
\left\langle m\left| {\hat{O}_{k}}\right| m\right\rangle =\exp (-\frac{\eta
^{2}}{2})\sum\limits_{p=0}^{m}\frac{(i\eta _{L})^{2p}m!}{p!\left( p+k\right)
!(m-p)!}.  \label{eq6}
\end{equation}
To obtain unitary time evolution of the system, we work in the basis $\left|
g,m,n\right\rangle ,$ $\left| e,m,n\right\rangle ,$ $\left|
g,m-1,n-1\right\rangle ,$ and $\left| e,m-1,n-1\right\rangle $ where $%
m,n=0,1,..,\infty $ denote the state of ionic vibrational motion and
quantized cavity field, respectively. The matrix representation of operator $%
\hat{H_{I}}$ in the chosen basis is 
\begin{equation}
H_{I}=\left[ 
\begin{array}{cccc}
0 & \hbar \Omega F_{m,m}^{L} & 0 & \hbar gF_{m,m-1}^{c}\sqrt{n} \\ 
\hbar \Omega F_{m,m}^{L} & 0 & 0 & 0 \\ 
0 & 0 & 0 & \hbar \Omega F_{m-1,m-1}^{L} \\ 
\hbar gF_{m,m-1}^{c}\sqrt{n} & 0 & \hbar \Omega F_{m-1,m-1}^{L} & 0
\end{array}
\right] ,  \label{eq7}
\end{equation}
where $F_{m,m}^{L}=\left\langle m\left| {\hat{O}_{0}}\right| m\right\rangle $%
, $F_{m,m-1}^{c}=\left\langle m\left| {\eta _{c}\hat{a}^{\dagger }\hat{O}%
_{1}^{c}}\right| m-1\right\rangle $. Working in the Lamb-Dicke regime that
is $\eta _{L}\ll 1$ and $\eta _{c}\ll 1$ , where $F_{m,m}^{L}\rightarrow 1$
for all $m$ and $F_{m,m-1}^{c}\rightarrow \eta _{c}\sqrt{m}$, we get to the
lowest order in $\eta _{L}$ and $\eta _{c},$ the matrix 
\begin{equation}
H_{I}^{LD}=\left[ 
\begin{array}{cccc}
0 & \hbar \Omega  & 0 & \hbar g\eta _{c}\sqrt{mn} \\ 
\hbar \Omega  & 0 & 0 & 0 \\ 
0 & 0 & 0 & \hbar \Omega  \\ 
\hbar g\eta _{c}\sqrt{mn} & 0 & \hbar \Omega  & 0
\end{array}
\right] .  \label{eq8}
\end{equation}
An analytic solution of time dependent Schr\"{o}dinger equation 
\begin{equation}
H_{I}^{LD}\Psi (t)=i\hbar \frac{d}{dt}\Psi (t)  \label{eq9}
\end{equation}
is easily found for a given initial state of the system. For the initial
states $\left| g,m-1,n-1\right\rangle $, and $\left| e,m-1,n-1\right\rangle $%
, defining $a=g\eta _{c}\sqrt{mn}$ and $\mu =\sqrt{a^{2}+\Omega ^{2}}$, we
obtain 
\begin{eqnarray}
\left| g,m-1,n-1\right\rangle  &\rightarrow &\left[ \frac{a}{\mu }\sin
(at)\sin (\mu t)+\cos (at)\cos (\mu t)\right] \left| g,m-1,n-1\right\rangle 
\nonumber \\
&&-i\left[ \frac{\Omega }{\mu }\cos (at)\sin (\mu t)\right] \left|
e,m-1,n-1\right\rangle   \nonumber \\
&&-\left[ \frac{\Omega }{\mu }\sin (at)\sin (\mu t)\right] \left|
g,m,n\right\rangle   \nonumber \\
&&+i\left[ \frac{a}{\mu }\cos (at)\sin (\mu t)-\sin (at)\cos (\mu t))\right]
\left| e,m,n\right\rangle   \label{eq10}
\end{eqnarray}
and 
\begin{eqnarray}
\left| e,m-1,n-1\right\rangle  &\rightarrow &\left[ \cos (at)\cos (\mu t)-%
\frac{a}{\mu }\sin (at)\sin (\mu t)\right] \left| e,m-1,n-1\right\rangle  
\nonumber \\
&&-i\left[ \frac{\Omega }{\mu }\cos (at)\sin (\mu t)\right] \left|
g,m-1,n-1\right\rangle   \nonumber \\
&&-\left[ \frac{\Omega }{\mu }\sin (at)\sin (\mu t)\right] \left|
e,m,n\right\rangle   \nonumber \\
&&-i\left[ \frac{a}{\mu }\cos (at)\sin (\mu t)+\sin (at)\cos (\mu t)\right]
\left| g,m,n\right\rangle   \label{eq10a}
\end{eqnarray}

From Eqs. (\ref{eq10} ,and \ref{eq10a}), for interaction time $t_{p}$ such
that $\mu t_{p}=p\pi $, $p=1,2,....$, we find the system in following three
qubit entangled states,

\begin{eqnarray}
\left| g,m-1,n-1\right\rangle  &\rightarrow &(-1)^{p}\left[ \cos
(at_{p})\left| g,m-1,n-1\right\rangle -i\sin (at_{p})\left|
e,m,n\right\rangle \right] ,  \label{11} \\
\left| e,m-1,n-1\right\rangle  &\rightarrow &(-1)^{p}\left[ \cos
(at_{p})\left| e,m-1,n-1\right\rangle -i\sin (at_{p})\left|
g,m,n\right\rangle \right] .  \label{11a}
\end{eqnarray}
When the initial state is $\left| g,m-1,n-1\right\rangle $, choosing $at_{p}=%
\frac{\pi }{4},$ we get maximally entangled tripartite two mode GHZ state,
with 
\begin{equation}
\left| g,m-1,n-1\right\rangle \rightarrow \frac{(-1)^{p}}{\sqrt{2}}\left(
\left| g,m-1,n-1\right\rangle -i\left| e,m,n\right\rangle \right) ,
\label{12}
\end{equation}
which for the choice $n=m=1$ corresponds to 
\begin{equation}
\left| g,0,0\right\rangle \rightarrow \frac{(-1)^{p}}{\sqrt{2}}\left( \left|
g,0,0\right\rangle -i\left| e,1,1\right\rangle \right) .  \label{12a}
\end{equation}
For this special case, where the resonator is initially in vacuum state, and
ion prepared in it's ground state with zero vibrational quanta, the shortest
operation time corresponds to the choice $p=1$. This implies that $\frac{\mu 
}{a}=4$ ; ${\mu }=\frac{4\Omega }{\sqrt{15}}$ giving $t_{1}\sim 0.34\mu s$,
using the value $\Omega =8.95MHz$ \cite{mund02}. To satisfy the condition $%
\frac{\mu }{a}=4$ we must fine tune the coupling constants and the cavity
Lamb-Dicke parameter such that $g=\frac{\Omega }{\eta _{c}\sqrt{15}}$. We
recall that if the ion is not placed exactly at the node, ion-quantized
field interaction is $\hbar g(\sigma _{+}+\sigma _{-})\left( \hat{b}%
^{\dagger }+\hat{b}\right) \sin \left[ \eta _{c}(\hat{a}^{\dagger }+\hat{a}%
)+\phi \right] .$ It is easily verified that the only change in final result
is that coupling constant $g$ is replaced by an effective coupling constant $%
g^{\prime }=g\cos \phi $. Adjustment of phase $\phi $ offers a mechanism for
the required fine tuning. The GHZ state can be detected by cavity-photon
measurement combined with atomic population inversion measurement.

We can verify that for initial states $\left| e,0,0\right\rangle $, $\left|
g,1,1\right\rangle $, and $\left| e,1,1\right\rangle $, after interaction
time $t_{1}$, the state of the system is 
\begin{eqnarray}
\left| e,0,0\right\rangle  &\Rightarrow &-\frac{1}{\sqrt{2}}\left( \left|
e,0,0\right\rangle -i\left| g,1,1\right\rangle \right) ,  \label{eq14} \\
\left| g,1,1\right\rangle  &\Rightarrow &-\frac{1}{\sqrt{2}}\left( \left|
g,1,1\right\rangle -i\left| e,0,0\right\rangle \right) ,  \label{eq15} \\
\left| e,1,1\right\rangle  &\Rightarrow &-\frac{1}{\sqrt{2}}\left( \left|
e,1,1\right\rangle -i\left| g,0,0\right\rangle \right) ,  \label{eq16}
\end{eqnarray}
where initial state is entangled with it's flipped state (having all three
qubit states rotated by $\pi $).

The merits of our proposal are simplicity of execution and short operation
time. In comparison, the implementation of a recent proposal \cite{feng02}
for preparation of GHZ state using two separate traps with a single ion
each, placed in a high-Q cavity, requires a qubit rotation, followed by two
CNOT gates on cavity and ionic vibrational states. Three qubit GHZ state
generated in our scheme is unique in that internal states and vibrational
states of massive ion are maximally entangled with photon which can serve as
an interface for quantum communication. It is important to note that
tripartite GHZ is generated without resorting to measurement and consequent
reduction. In case the purpose is using maximally entangled state for
quantum computation, teleportation or cloning, state reduction is not a
desirable feature.

S.S.S thanks Unicamp for hospitality.

\end{document}